\documentclass{elsart}
\usepackage{epsfig}

\newcommand{\asc}[1]%
             {$\langle${#1}$\rangle$}

\begin{document}

\begin{frontmatter}
\title{Impact of Varying Atmospheric Profiles on Extensive Air Shower Observation: \\
 - Fluorescence Light Emission and Energy Reconstruction - }

\author[Uni]{B.
Keilhauer}$^,$\renewcommand{\thefootnote}{\fnsymbol{footnote}}\footnote{Corresponding
author. \em{E-mail-Address:} bianca.keilhauer@ik.fzk.de}, 
\author[Uni,FZK]{J. Bl\"umer},
\author[FZK]{R. Engel},
\author[FZK]{H.O. Klages}

\address[Uni]{Universit\"at Karlsruhe, Institut f\"ur Experimentelle Kernphysik, Postfach
3640, 76021 Karlsruhe, Germany}
\address[FZK]{Forschungszentrum Karlsruhe, Institut f\"ur Kernphysik, Postfach 3640, 76021
Karlsruhe, Germany}

\begin{abstract}
Several experiments measure the fluorescence light produced by extensive air showers in the
atmosphere. This light is converted into a longitudinal shower profile from which information
on the primary energy and composition is derived. 
The fluorescence yield, as the conversion factor between light profile measured by EAS experiments 
and physical interpretation of showers, has been measured in several laboratory experiments. 
The results, however, differ considerably. In this article, a model calculation of the fluorescence emission 
from relevant band systems of nitrogen in dependence on wavelength and atmospheric conditions is presented.
Different calculations are compared to each other in combination with varying input parameters.
The predictions are compared with measurements and the altitude-dependence of the
fluorescence yield is discussed in detail.
 \newline
\begin{small}
\end{small}

\end{abstract}
\end{frontmatter}

\section{Introduction\label{sec:intro}}

This is the second article of a series of investigations of the importance of 
atmospheric properties for the reconstruction of extensive air showers (EAS). 
The first article~\cite{keilhauer04} describes the effect of changing atmospheric density 
profiles on the longitudinal EAS development. In particular, if the fluorescence 
technique is applied, the conversion of atmospheric depth, as used in shower simulations, to geometrical
altitude, as reconstructed from fluorescence measurements, and vice versa is very important
for primary cosmic ray mass reconstruction.

This article addresses the fluorescence light emission of EAS which is used for the determination of the 
total energy of EAS. In several air shower experiments, for 
example, HiRes~\cite{hires}, the Pierre Auger Observatory~\cite{PAO1,PAO2,PAO3}, and Telescope
Array~\cite{TA1,TA2}, the fluorescence technique is employed for detecting EAS. Measuring the 
fluorescence light that nitrogen molecules emit after being excited by charged particles 
traversing the atmosphere is the most direct method of detecting the longitudinal shower 
profile. For the event reconstruction procedures of these air shower experiments, the knowledge of 
the fluorescence
yield $FY_\lambda$ and its dependence on atmospheric conditions are crucial parameters.

The Pierre Auger Observatory is up to now the only existing EAS experiment which applies hybrid
detection techniques. The secondary particles of an EAS are measured at ground and simultaneously
the fluorescence light of the longitudinal shower development is detected with telescopes. For extracting
a cosmic ray spectrum from the data, the events detected with ground detectors are analyzed while the
energy calibration is deduced from fluorescence detector events and the correlation of these two
types of events is derived from hybrid events~\cite{augerspec_ICRC}. This cosmic ray spectrum and 
its comparison with spectra published by other experiments (AGASA~\cite{agasa} and
HiRes~\cite{hires_spec}, for a recent review see~\cite{engel}) reveal
that the fluorescence yield might be a crucial parameter for the energy reconstruction of air showers.
For the conversion of detected fluorescence light to energy deposited in the atmosphere by EAS and
finally to the total energy of the primary particle, not only the total fluorescence yield
in the detected wavelength ($\lambda$) region is important but also the spectral
distribution. For example, the emitted light
suffers \emph{Rayleigh} scattering while traversing the atmosphere towards the telescopes. Since
the scattering cross section has a $\lambda^{-4}$ dependence, the long-wavelength part of the
fluorescence spectrum has a higher transmission than the short-wavelength region. 
 
The outline of this paper is as follows. In Section~\ref{sec.theory}, the fluorescence emission in air is 
reviewed and an analytical model (Sec.~\ref{sec.mathdesc}) for calculating the fluorescence emission 
in dependence on wavelength and atmospheric conditions is described. Particularly, the 
band systems of nitrogen contributing mainly to the fluorescence light 
emission are discussed. A compilation of several 
parameters from different authors used in calculations of this paper is given in Sec.~\ref{sec.param}.
In Sec.~\ref{sec.comp} a detailed comparison of measurements with the calculations is done. The aim 
is to combine the laboratory measurements with our current understanding of the
fluorescence emission processes in the atmosphere, to show possible sources of
uncertainty, and to provide an easy way of accounting for varying atmospheric
conditions. The dependence on these atmospheric conditions is explicitly presented 
in Sec.~\ref{sec.althumdep}. 

\section{Model for Fluorescence Light Emission in Air\label{sec.theory}}

The most numerous charged particles in an EAS are electrons and positrons. Their energy deposit
in air by ionization and excitation of air molecules gives rise to fluorescence light
emission. In the wavelength region between 300 and 400~nm, the band systems with the
strongest emissions are found. Therefore, all EAS experiments using the fluorescence technique apply UV filters
with largest transmittance between roughly 310 and 400~nm. The residual wavelengths are cut in favor of
reducing the night sky background. 

The major components of the atmosphere are 78.08\% N$_2$, 20.95\% O$_2$, and 0.93\% Ar per
volume. All three constituent parts influence the emission of fluorescence light,
however, with strongly differing importance.

The main fluorescence light is emitted by two electronic states of N$_2$, these are the
second-positive (2P) band system, $C^3\Pi_u$ - $B^3\Pi_g$, and the first-negative (1N) system 
of N$^+_2$, $B^2\Sigma^+_u$ - $X^2\Sigma^+_g$.
Each band within a system belongs to a transition from a vibrational level of the upper state 
$\nu^\prime$ to a vibrational level of the lower state $\nu^{\prime\prime}$~\cite{haken}.
The band structure is caused by the rotational levels and consists of a large number of spectral 
lines which are very close to each other~\cite{herzberg}. These lines will not be distinguished in this work.
The bands usually have a \emph{band head} at one end where the intensity falls off
suddenly. In the following, the wavelength given for each band belongs to the position of the 
corresponding band head.

For both, absorption and emission, the 
\emph{Franck-Condon principle} applies in good approximation. Three processes of excitation
of N$_2$ can be distinguished:
\begin{itemize}
\item Direct excitation: The energy deposited in air excites nitrogen molecules proportional
to an energy-dependent cross section $\sigma_{\nu^\prime}(E)$. This process mainly acts on
the N$^+_2$ 1N system
\begin{eqnarray}
{\rm N}_2 + e \rightarrow {\rm N}^{+\ast}_2 + e + e.
\end{eqnarray}
\item Excitation via secondary electrons: High energy particles in EAS ionize N$_2$ producing
several lower energy secondary electrons. These $e^-$ are able to excite also the N$_2$ 2P system
with a resultant spin change
\begin{eqnarray}
{\rm N}_2 + e(\uparrow) \rightarrow {\rm N}^\ast_2(C^3\Pi_u) + e(\downarrow).
\end{eqnarray}
In addition, the 2P system can also be excited by recombination
\begin{eqnarray}
{\rm N}_2^+ + e \rightarrow {\rm N}^\ast_2(C^3\Pi_u).
\end{eqnarray}
\item Via \emph{Auger electrons}: Some ionization processes will 
release K-electrons leading to the emission of Auger electrons. These are on their part
again able to excite the N$_2$ molecules. However, it must be emphasized that the cross section for
this K-shell ionization is much lower than those for the processes mentioned above. The cross sections 
for N$_2$ excitation and for the ionization to N$_2^+$ are of the order of 
10$^{-21}$~m$^2$~\cite{fons1994,itikawa} and for the K-shell ionization of the order of 
10$^{-23}$~m$^2$~\cite{santos}.
\end{itemize}

In air, the optical emission of the prompt radiative return from the upper states of the 
2P and 1N system of nitrogen will be affected by some competing processes. 
The most important process is collisional quenching. Excited nitrogen molecules might collide with other
molecules in air before the de-excitation via fluorescence light emission happens.

Argon can be excited by the reaction $e + {\rm Ar} \rightarrow {\rm Ar}^\ast$, where the 
excitation cross section is largest for Ar($^3P_2$) \cite{bennett}. This process is 
followed by ${\rm Ar}^\ast + {\rm N}_2 \rightarrow {\rm Ar} + {\rm N}_2^\ast(C^3\Pi_u)$. The 
energy is mainly transferred from argon to nitrogen via secondary electrons rather than direct
collisions \cite{grun}. The excited state $C^3\Pi_u$ is the upper level of the second-positive 
(2P) system of N$_2$ which radiates photons mainly in the wavelength region between 300 and 
400~nm. The lower state is $B^3\Pi_g$. This increase of the emission competes, however, with a 
higher quenching rate, which leads to non-radiative de-excitation, due to additional collision 
partners in the form of argon atoms in air. The net effect of argon is expected to be less than 
1\% contribution to the fluorescence light \cite{bunner}. However, argon emits also directly 
fluorescence light at around 310~nm~\cite{ulrich2}. This transition, $A^2\Sigma^+ - X^2\Pi$, 
has been investigated in argon water-vapor mixtures and the highest intensity has been found 
for very low argon pressure and 0.06~hPa water vapor. For EAS experiments, this contributions
will be of minor importance, too. 

The UV-fluorescence light emission from O$_2$ is negligible \cite{nicholls}. The contribution
in the relevant wavelength region stems from O$^+_2 A^2\Pi_u - X^2\Pi_g$ transmissions.
However, already the \emph{Einstein coefficients}\footnote{radiative transition probabilities}
are reduced in average by a factor of about 30 compared to the emissions of the 2P system of 
N$_2$ \cite{gilmore}. The emissions of atomic oxygen start at 395~nm and go up to 845~nm \cite{nicholls}. 
These bands are of no importance for EAS experiments.

Generally, it is assumed that the fluorescence light is proportional to the energy deposit of 
an EAS. The contribution of electrons and positrons to the energy deposit 
according to the initial kinetic energy distribution in an air shower has been studied 
elsewhere \cite{risse2004}. Only 10\% of the energy deposit stems from particles with 
energies less than 0.1~MeV.
Particles with energies between 0.1 and 10~MeV contribute 35\%, between 10 and 100~MeV also 35\%, 
and between 100 and 1000~MeV 17\%. The remaining 3\% originate from particles of energy
above 1000~MeV. Several experiments have recently started to measure the 
proportionality of fluorescence yield to energy deposit. So far none of the results on the energy
dependence is absolutely 
calibrated, thus a direct comparison is difficult. First relative results seem to confirm the expected
correlation in different energy regions, see e.g.~at about 1~MeV Nagano et al.~\cite{nagano2003}, 
between 50 and 420~MeV Bohacova et al.~\cite{bohacova}, and at 28.5~GeV Belz et al.~\cite{belz}.

Depending on their initial energy, EAS particles produce secondary electrons of various lower 
energies. These can excite the N$_2$ but they may also suffer an 
\emph{attachment process}: if, on their way from the production site to the N$_2$ molecules, the 
secondary electrons encounter a strong electronegative pollutant (e.g.~O$_2$, H$_2$O, CO$_2$, H$_2$, 
Xe, CH$_4$ which are trace gases in the atmosphere), they are attached to this pollutants and 
cannot excite the N$_2$ molecules anymore \cite{grun}. This process is beyond the scope of this paper.

\subsection{Mathematical Description\label{sec.mathdesc}}

The existing results of fluorescence yield measurements show quite large differences. Furthermore, the
data have to be applied to air shower reconstruction procedures. While secondary particles of EAS 
traverse from high to low altitudes, they encounter continuously changing atmospheric conditions. Additionally,
the atmospheric conditions vary from day to day with the largest differences between the seasons summer and
winter at the sites of all existing air shower experiments. The aim of the calculations shown here 
is to cross-check the laboratory measurements with the current understanding of the
processes in the atmosphere, show possible sources of  
uncertainties, and provide an easy way to account for varying atmospheric conditions in
EAS measurements.

The quantum efficiency of fluorescence can be defined as 
\begin{eqnarray}
\frac{\textrm{rate of de-excitation via radiation}}{\textrm{total rate of de-excitation}} =
\frac{\tau_c}{\tau_0 + \tau_c}{\rm  ,}
\end{eqnarray}
where the rate of de-excitation is proportional to the reciprocal of the life time. The mean life time
of the radiative transition to any lower state is $\tau_0$ and to collisional quenching $\tau_c$. The 
collisional quenching can be described by kinetic gas theory. The molecules, in the case of air, move 
with velocities following the Maxwell-Boltzmann distribution which is strongly correlated with
gas temperature. As a good approximation, the collision rate depends on the mean velocity of molecules 
$\overline{v}$ = $\sqrt{\frac{8kT}{\pi M}}$. The resulting mean life time due to collisional quenching
is the ratio of the mean free path, in this case for molecules of one type moving with  
roughly the same velocity, and the mean velocity:
\begin{eqnarray}
\tau_c = (\sqrt{2}\cdot \rho_n\cdot \sigma_{{\rm NN}}\cdot\overline{v})^{-1} = \sqrt{(\pi
M/kT)}\cdot (4\rho_n\cdot \sigma_{{\rm NN}})^{-1},
\end{eqnarray}
where $\rho_n$ is the particle number density, $\sigma_{{\rm NN}}$ the collisional cross section 
between nitrogen molecules, 
$T$ the temperature, $k$ the Boltzmann constant, and $M$ the molecular
mass. Now the pressure dependent fluorescence efficiency can be written as 
\begin{eqnarray}
\varepsilon_\lambda(p,T)&=&\frac{\varepsilon_\lambda^0}{1+(p/p_{\nu^\prime}^\prime(T))}
 = \frac{n\cdot E_\gamma}{E_{dep}},
\end{eqnarray}
with $\varepsilon_\lambda^0$ being the fluorescence efficiency at wavelength
$\lambda$ without collisional quenching, $n$ denoting the number of photons, $E_\gamma$ the energy
of a single photon with the corresponding wavelength, $E_{dep}$ the deposited energy in the observed 
medium, and $p/p^\prime_{\nu^\prime}$ =
$\tau_{0,{\nu^\prime}}/\tau_{c,{\nu^\prime}}$. The pressure $p$ is that of the observed medium (e.g.~air),
$p^\prime_{\nu^\prime}$ is a reference pressure at which $\tau_0$ is equal to
$\tau_c$. $\tau_{0,\nu^\prime}$ and 
$\tau_{c,\nu^\prime}$ are the mean life times for excitation level $\nu^\prime$. Applying actual 
atmospheric conditions, 
with air taken to be a two-component gas, the relation between $p$ and $p^\prime_{\nu^\prime}$ can 
be written as
\begin{eqnarray}
\frac{p}{p^\prime_{\nu^\prime}}&=& \tau_{0,\nu^\prime}\cdot
\biggl(\frac{1}{\tau_{{\rm NN},\nu^\prime}(\sigma_{{\rm NN},\nu^\prime})}
+\frac{1}{\tau_{{\rm NO},\nu^\prime}(\sigma_{{\rm NO},\nu^\prime})}\biggr) \label{eq.ppprime} \\
&=& \frac{\tau_{0,\nu^\prime}
p_{{\rm air}}\cdot N_A}{R\cdot T}\cdot\sqrt{\frac{kTN_A}{\pi}}
\cdot\biggl(4\cdot C_v({\rm N}_2)\cdot\sigma_{{\rm NN},\nu^\prime}
\sqrt{\frac{1}{M_{m,{\rm N}}}} \label{eq.ppprimelong}
\end{eqnarray}
\vspace{-34pt}
\begin{eqnarray*} 
&~&~~~~~~~~~~~~~~~~~~~~~~~~~~~+~2\cdot
C_v({\rm O}_2)\cdot\sigma_{{\rm NO},\nu^\prime}\sqrt{2(\frac{1}{M_{m,{\rm N}}}
+\frac{1}{M_{m,{\rm O}}})}\biggr), 
\end{eqnarray*}
with Avogadro's number $N_A$, the masses per mole for nitrogen $M_{m,{\rm N}}$ and oxygen
$M_{m,{\rm O}}$, the universal gas constant $R$, the cross sections for collisional 
de-excitation for nitrogen-nitrogen $\sigma_{{\rm NN},\nu^\prime}$ and nitrogen-oxygen 
$\sigma_{{\rm NO},\nu^\prime}$, and the fractional part per volume $C_v$ of the two gas components. 

\subsection{Input Parameters\label{sec.param}}

To estimate the fluorescence efficiency with these equations, several parameters have to
be obtained from measurements and/or calculations. 
Most important is the fluorescence efficiency without collisional quenching $\varepsilon_\lambda^0$. An early
compilation of elder measurements performed by Bunner~\cite{bunner} provides these values for 18 band systems 
of the 2P and for 1 band system of the 1N state of nitrogen in the wavelength region between 300 and 400~nm. 
In a more recent publication by Gilmore et al.~\cite{gilmore}, the Einstein coefficients 
$A_{\nu^\prime\nu^{\prime\prime}}$ of the 2P 
nitrogen state for the transitions from $\nu^\prime = 0 \ldots 4$ to $\nu^{\prime\prime} = 0 \ldots 21$ and
the radiative life times $\tau_{0,\nu^\prime}$ for $\nu^\prime = 0 \ldots 4$ are given. For the 1N 
nitrogen state, the Einstein
coefficients $A_{\nu^\prime\nu^{\prime\prime}}$ for the transitions from $\nu^\prime = 0 \ldots 10$ 
to $\nu^{\prime\prime} = 0 \ldots 21$ and
the radiative life times $\tau_{0,\nu^\prime}$ for $\nu^\prime = 0 \ldots 10$ are listed. 
The intensity of a transition could be calculated by $N_{\nu^\prime \rightarrow \nu^{\prime\prime}} 
= \tau_{0,\nu^\prime} \cdot A_{\nu^\prime\nu^{\prime\prime}}
\cdot N^\ast_{\nu^\prime}$, where $N^\ast_{\nu^\prime}$ is the number of excited states. Since this number 
is unknown, a relative fluorescence efficiency is calculated by multiplying the Einstein coefficients with 
the radiative life times and a \emph{relative apparent excitation cross section} $Q_{app}$ ~\cite{fons1996}. 
The \emph{apparent excitation cross sections} are the sum of the optical emission cross sections over 
$\nu^{\prime\prime}$ for a given $\nu^\prime$. The excitation
cross sections are called apparent, because they represent the sum of direct excitation
and cascades from higher levels down to that particular one.
The $Q_{app}$ values are derived from the \emph{apparent excitation cross sections} by normalizing the value
for $\nu^\prime = 0$ to unity of the electronic state. Values are given in the publication
by Fons et al.~\cite{fons1996} for the 2P band system of nitrogen for $\nu^\prime = 0
\ldots 4$ and for the 1N band system for $\nu^\prime = 0\ldots 3$ in Stanton and St.~John~\cite{StSt}. 

The relative fluorescence efficiency can then be normalized for each electronic state to,
e.~g., the most prominent band. For the 2P system of
nitrogen, it is the (0-0) band with a wavelength of 337.1~nm, and for the 1N system it is
again the (0-0) band with the corresponding wavelength of 391.4~nm. In our calculations, the
efficiency values for these wavelengths as given in Bunner~\cite{bunner} are chosen. 

In the following the fluorescence efficiencies are calculated for the bands of the 2P system.
They dominate the fluorescence light emission in the atmosphere due to excitations caused by EAS.
The fluorescence efficiencies labeled with $\varepsilon_{\lambda,G.-F.}$ are obtained by applying
the Einstein coefficients given by Gilmore et al.~and the relative apparent excitation cross
sections from Fons et al. The values used within this paper are listed in 
Tab.~\ref{tab:fleff}.
\begin{table}[htbp]
\caption{Constants for the fluorescence efficiency $\varepsilon_\lambda^0$ of the 2P
system of nitrogen.\label{tab:fleff}}
\begin{center}
\begin{tabular}{|c|c||c||c|c|c||c|}\hline
Band & Wave- &  & \multicolumn{2}{c|}{Gilmore et al.~\cite{gilmore}} &
Fons &  \\
  & length & $\varepsilon_{\lambda,Bunner}^0$ & $A_{\nu^\prime\nu^{\prime\prime}}$ &  $\tau_{0,\nu^\prime}$ 
& et al.~\cite{fons1996}  & $\varepsilon_{\lambda,G.-F.}^0$  \\
($\nu^\prime - \nu^{\prime\prime}$) & $\lambda$ (nm) & (\%) & (1/s) & (s) & $Q_{app}$ & (\%) \\
 \hline
2P (0-0) & 337.1 & .082 & 1.31E7 & 3.71E-8 & 1 & .082 \\
2P (0-1) & 357.7 & .0615 & 8.84E6 & 3.71E-8 & 1 & .0553 \\
2P (0-2) & 380.5 & .0213 & 3.56E6 & 3.71E-8 & 1 & .0223 \\ \hline
2P (1-0) & 315.9 & .050 & 1.19E7 & 3.75E-8 & 0.7 & .0527 \\
2P (1-1) & 333.9 & .0041 & 5.87E5 & 3.75E-8 & 0.7 & .0026 \\
2P (1-2) & 353.7 & .029 & 5.54E6 & 3.75E-8 & 0.7 & .0245 \\
2P (1-3) & 375.5 & .0271 & 4.93E6 & 3.75E-8 & 0.7& .0218 \\
2P (1-4) & 399.8 & .016 & 2.43E6 & 3.75E-8 & 0.7 & .0108 \\ \hline 	
2P (2-1) & 313.6 & .029 & 1.01E7 & 3.81E-8 & 0.26 & .0169 \\
2P (2-2) & 330.9 & .002 & 7.99E5 & 3.81E-8 & 0.26 & .0013 \\
2P (2-3) & 350.0 & .004 & 1.71E6 & 3.81E-8 & 0.26 & .0029 \\
2P (2-4) & 371.0 & .010 & 4.04E6 & 3.81E-8 & 0.26 & .0068 \\
2P (2-5) & 394.3 & .0064 & 3.14E6 & 3.81E-8 & 0.26 & .0052 \\ \hline
2P (3-2) & 311.7 & .005 & 5.94E6 & 3.90E-8 & 0.081 & .0032 \\
2P (3-3) & 328.5 & .0154 & 2.85E6 & 3.90E-8 & 0.081 & .0015 \\
2P (3-4) & 346.9 & .0063 & 1.15E5 & 3.90E-8 & 0.081 & .0001 \\
2P (3-5) & 367.2 & .0046 & 2.35E6 & 3.90E-8 & 0.081 & .0013 \\
2P (3-6) & 389.5 & .003 & 3.00E6 & 3.90E-8 & 0.081 & .0016 \\ \hline 
2P (4-3) & 310.4 & - & 3.02E6 & 4.04E-8 & 0.041 & .0008 \\
2P (4-4) & 326.8 & - & 3.71E6 & 4.04E-8 & 0.041 & .0010 \\
2P (4-5) & 344.6 & - & 1.24E5 & 4.04E-8 & 0.041 & .0000 \\
2P (4-6) & 364.2 & - & 9.98E5 & 4.04E-8 & 0.041 & .0003 \\
2P (4-7) & 385.8 & - & 2.33E6 & 4.04E-8 & 0.041 & .0007 \\ \hline
\end{tabular}
\end{center}
\vspace{40pt}
\end{table}

Further parameters in the upper equations are the deactivation constants which are the radiative life time 
$\tau_{0,\nu^\prime}$ and the collisional cross section between nitrogen and nitrogen molecules 
$\sigma_{NN,\nu^\prime}$ and between nitrogen and oxygen molecules $\sigma_{NO,\nu^\prime}$. 
The values for $\tau_{0,\nu^\prime}$ obtained by Gilmore et al.~are listed 
in Tab.~\ref{tab:fleff} and \ref{tab:fleff_1N}. Bunner provides collisional cross sections and
radiative life times for the most prominent band systems of nitrogen, see Tab.~\ref{tab:deact}. Recent
measurements by Morozov et al.~\cite{ulrich} were performed for the 2P $\nu^\prime = 0,1$
band systems, see also Tab.~\ref{tab:deact}. In further calculations presented in this
article, labeled with \emph{Morozov}, the values from Bunner are replaced by the newer
data by Morozov et al.~where available. 
\begin{table}[tbp]
\caption{Deactivation constants for air in the lower atmosphere.\label{tab:deact}}
\begin{minipage}{\linewidth}
\renewcommand{\thefootnote}{\thempfootnote}
\begin{center}
\begin{tabular}{|r||c|c|c||c|c|c|}\hline
 & \multicolumn{3}{c||}{Bunner~\cite{bunner}} & \multicolumn{3}{c|}{Morozov et al.~\cite{ulrich}} \\
 & $\sigma_{{\rm NO}}$ & $\sigma_{{\rm NN}}$ & $\tau_0$ & $\sigma_{{\rm NN}}$ & $\sigma_{{\rm N}vapor}$ & $\tau_0$ \\
 & (m$^2$) & (m$^2$) & (s) & (m$^2$) & (m$^2$) & (s) \\ \hline  \hline
1N $\nu$ = 0 & 13E-19 & 4.37E-19 & 6.58E-8 & - & - & - \\
2P $\nu$ = 0 & 2.1E-19 & 1.0E-20 & 4.45E-8 & 1.82E-20 & 8.53E-19 & 4.17E-8 \\
   $\nu$ = 1 & 5.0E-19 \footnote{This value is determined by the given results of \cite{bunner} and not given in the original publication.} & 3.5E-20 & 4.93E-8 & 3.77E-20 & 8.04E-19 & 4.17E-8 \\
   $\nu$ = 2 & 7.0E-19 \footnotemark[\value{footnote}] & 8.8E-20 & 4.45E-8 & - & - & - \\
   $\nu$ = 3 & 8.0E-19 \footnotemark[\value{footnote}] & 1.2E-19 & 6.65E-8 & - & - & - \\ \hline
\end{tabular}
\end{center}
\end{minipage}
\end{table}

No collisional cross sections are available for the first-negative system of nitrogen for
higher excitation levels than the (0-0) band. For an estimation of higher level contributions, one
can assume that the collisional cross section between the ionized nitrogen molecule and
further nitrogen and oxygen molecules are equal to that of the (0-0) transition. Since
the collisional cross section usually increases for
higher excitations, this is an assumption which gives an upper bound on the fluorescence light
from these levels. The relative fluorescence yields for the 1N bands are estimated with the same 
method as already used for 2P, however, applying the parameters given in Gilmore et al.~and
Stanton and St.~John (see $\varepsilon_{\lambda,G.-St.}^0$ in
Tab.~\ref{tab:fleff_1N}). Even under these conditions, the 1N (1-0) and 1N (1-1)
transitions add up to less than 1\% to the total fluorescence yield between 300 and
400~nm at sea level in the US Standard Atmosphere. Higher excitations give even smaller 
contributions. Therefore, we will only consider the (0-0) transition of the 1N band and
neglect the higher excitations.
\begin{table}[htbp]
\caption{Constants for the fluorescence efficiency $\varepsilon_\lambda^0$ of the 1N
system of nitrogen.\label{tab:fleff_1N}}
\begin{center}
\begin{tabular}{|c|c||c||c|c|c||c|}\hline
Band & Wave- &  & \multicolumn{2}{c|}{Gilmore et al.~\cite{gilmore}} &
Stanton \& &  \\
  & length & $\varepsilon_{\lambda,Bunner}^0$ & $A_{\nu^\prime\nu^{\prime\prime}}$ &  $\tau_{0,\nu^\prime}$ 
& St.~John~\cite{StSt}  & $\varepsilon_{\lambda,G.-St.}^0$  \\
($\nu^\prime - \nu^{\prime\prime}$) & $\lambda$ (nm) & (\%) & (1/s) & (s) & $Q_{app}$ & (\%) \\
 \hline
1N (0-0) & 391.4 & .33 & 1.14E7 & 6.23E-8 & 1 & 0.33 \\ \hline
1N (1-0) & 358.0 & - & 5.76E6 & 6.20E-8 & 0.117 & 0.0193 \\
1N (1-1) & 388.2 & - & 4.03E6 & 6.20E-8 & 0.117 & 0.0135 \\ \hline
1N (2-0) & 330.5 & - & 9.02E5 & 6.19E-8 & 0.009 & 0.0002 \\
1N (2-1) & 356.1 & - & 7.88E6 & 6.19E-8 & 0.009 & 0.0021 \\
1N (2-2) & 385.5 & - & 9.27E5 & 6.19E-8 & 0.009 & 0.0003 \\ \hline
1N (3-1) & 329.6 & - & 2.08E6 & 6.23E-8 & 0.004 & 0.0002 \\
1N (3-2) & 354.6 & - & 8.09E6 & 6.23E-8 & 0.004 & 0.0009 \\ \hline
\end{tabular}
\end{center}
\vspace{40pt}
\end{table}

The calculation introduced in Sec.~\ref{sec.mathdesc} together with the
$\varepsilon_{\lambda,Bunner}^0$ and 
the deactivation constants from \emph{Morozov} is used as reference model in this
paper. It benefits from the completeness 
of the Bunner data and from the accuracy of the measurements from Morozov et al.

\section{Comparison with Measurements\label{sec.comp}}

\begin{figure}[htbp]
\centering\epsfig{file=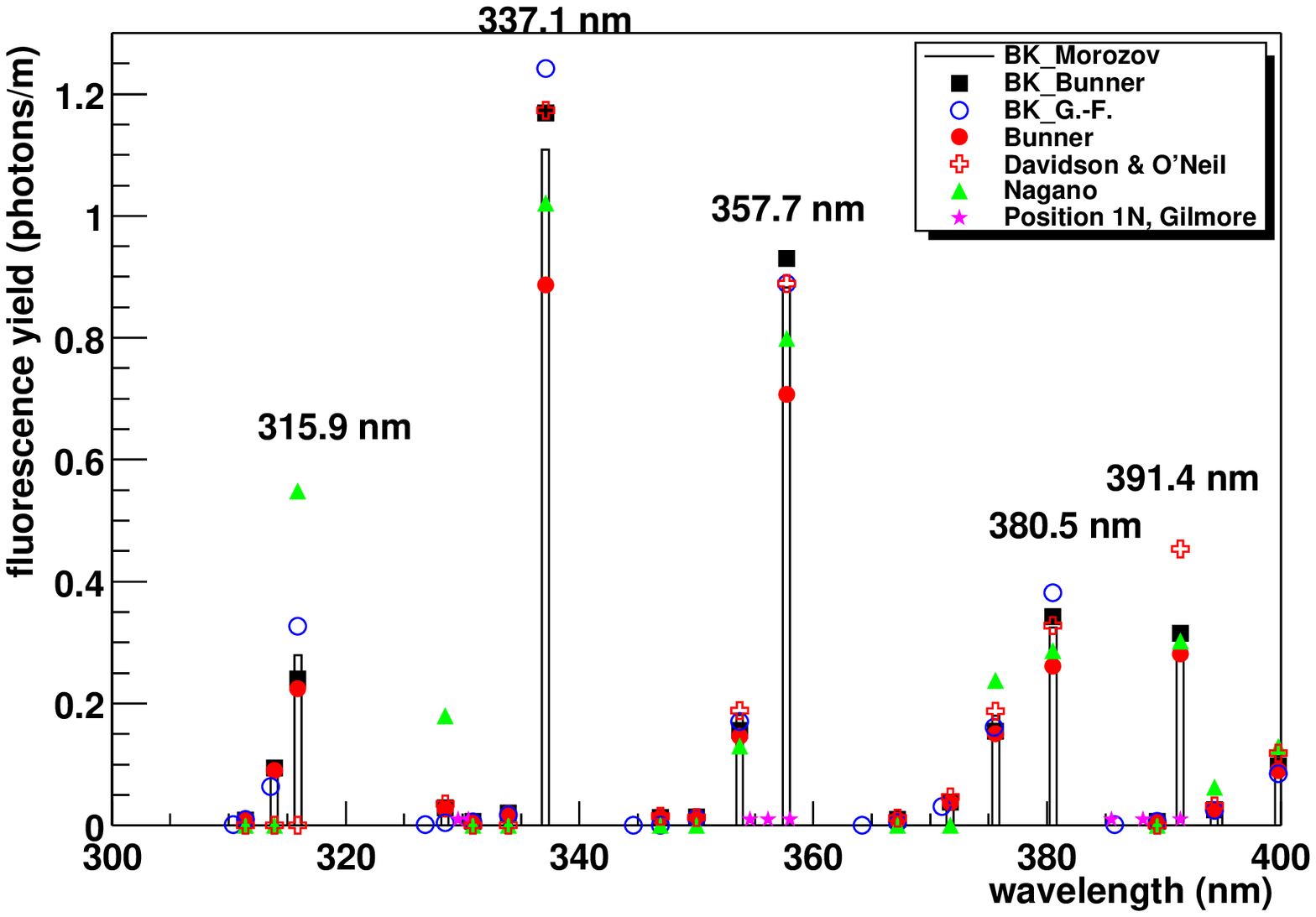,width=.9\linewidth}
\caption{Fluorescence yield spectra of several calculations and measurements for 0.85~MeV electrons as 
exciting particles. The bars indicate the preferred calculation presented in this article. 
All calculations are labeled with ``BK\_\emph{name}'', where
\emph{name} indicates the authors of the input parameters. Pink stars indicate the positions of possible 
contributions of the 1N system beyond the 1N (0-0) transition.\label{fig:spec_yield}}
\end{figure}

Wavelength-dependent results of fluorescence yield measurements have been provided in three
publications~\cite{bunner,nagano,DO}. Bunner lists several intermediate values: $\varepsilon_\lambda^0$, 
$\varepsilon_\lambda^{s.l.}(p,T)$ in \%, and the fluorescence efficiency $\varepsilon_{E_{dep}}^{s.l.}$ 
in units of photons/MeV of deposited energy which is $\varepsilon_\lambda^{s.l.}(p,T)
\cdot (\lambda/hc)$ at sea level ($s.l.$). The values for  
$\varepsilon_\lambda^{s.l.}(p,T)$ and $\varepsilon_{E_{dep}}^{s.l.}$ given explicitly in~\cite{bunner} are
not reproduced by the calculations shown here, see Table~\ref{tab:yield}. Possible reasons are 
rounding uncertainties by Bunner or the use of deviating numbers for variables concerning air conditions. 
Davidson and O'Neil~\cite{DO} list results for $\varepsilon_\lambda^{s.l.}(p,T)$ for wavelengths above 
320~nm. It should be mentioned that the results in~\cite{DO} are given for $p$~=~800~hPa. 
The increase of the total fluorescence yield between 300
and 400~nm from sea level with $p$~=~1013~hPa to approximately 2~km~a.s.l.~with $p$~=~800~hPa amounts 
to about 2\%. Nagano et al.~report directly the values for $FY_\lambda$ at sea level
for 0.85~MeV electrons~\cite{nagano}, however, only 10 contributing emission bands are listed. 
For comparing the results of all authors, 0.85~MeV electrons are chosen as exciting particles, so the 
ionization energy deposit is $dE/dX$ = 0.1677 MeV/kg$\cdot$m$^{-2}$~\cite{nagano2003}. It is assumed that 
the fluorescence yield is proportional to the energy deposit as discussed in Sec.~\ref{sec.theory}. Air is 
taken to be a composition of 78.8\% N$_2$ and 21.1\% O$_2$~per volume \cite{nagano2003}. The resulting 
fluorescence yield can be written as
\begin{eqnarray}
FY_\lambda ~=~
\varepsilon_\lambda(p,T)\cdot\frac{\lambda}{hc}\cdot\frac{dE}{dX}\cdot\rho_{air}~\biggl[\frac{{\rm photons}}{{\rm m}}\biggr]. \label{eq.FY}
\end{eqnarray}
A comparison of the obtained $FY_\lambda$ values at sea level in the US Standard 
Atmosphere (US-StdA)~\cite{US-StdA1976,keilhauer04} is shown in Table~\ref{tab:yield}
\begin{table}[htbp]
\caption{Fluorescence yield at sea level in the US Standard Atmosphere. For comparing the results of all 
authors, 0.85~MeV electrons are chosen as exciting particles, so the ionization energy deposit is 
0.1677 MeV/kg$\cdot {\rm m}^{-2}$. See text for details.\label{tab:yield}}
\renewcommand{\thefootnote}{\thempfootnote}
\begin{minipage}{\linewidth}
\begin{center}
\begin{tabular}{|c||c|c|c||c|c|c|}\hline
 & \multicolumn{3}{c||}{$FY_\lambda^{s.l.}~(\frac{{\rm photons}}{{\rm m}})$ measurements from} 
& \multicolumn{3}{c|}{$FY_\lambda^{s.l.}~(\frac{{\rm photons}}{{\rm m}})$ calculations with} \\
Wave- & Bunner & Davidson \& & Nagano & $\varepsilon_{\lambda,Bunner}^0$, & $\varepsilon_{\lambda,Bunner}^0$, 
& $\varepsilon_{\lambda,G.-F.}^0$, \\
length & \cite{bunner} & O'Neil~\cite{DO} & et al.~\cite{nagano}  & Table~\ref{tab:deact}~- & Table~\ref{tab:deact}~- & 
$\sigma_{Nx,Morozov}$, \\
$\lambda$ (nm) &  & & & Bunner & Morozov & $\tau_{0,Gilmore}$ \\ \hline \hline
310.4 & -\footnote{This transition has not been measured.} & -\footnote{Only measurements above 320~nm.} & -\addtocounter{mpfootnote}{-1}\footnotemark[\value{mpfootnote}] & - & - & 0.001 \\
311.7 & 0.008 & -\addtocounter{mpfootnote}{+1}\footnotemark[\value{mpfootnote}] & -\addtocounter{mpfootnote}{-1}\footnotemark[\value{mpfootnote}] & 0.009 & 0.009 & 0.010 \\
313.6 & 0.090 & -\addtocounter{mpfootnote}{+1}\footnotemark[\value{mpfootnote}] & -\addtocounter{mpfootnote}{-1}\footnotemark[\value{mpfootnote}] & 0.094 & 0.094 & 0.064 \\
315.9 & 0.224 & -\addtocounter{mpfootnote}{+1}\footnotemark[\value{mpfootnote}] & 0.549 & 0.240 & 0.279 & 0.326 \\
326.8 & -\addtocounter{mpfootnote}{-1}\footnotemark[\value{mpfootnote}] & -\footnotemark[\value{mpfootnote}] & -\footnotemark[\value{mpfootnote}] & - & - & 0.002 \\
328.5 & 0.027 & 0.035 & 0.180 & 0.029 & 0.029 & 0.005 \\
330.9 & 0.007 & -\footnotemark[\value{mpfootnote}] & -\footnotemark[\value{mpfootnote}] & 0.007 & 0.007 & 0.005 \\
333.9 & 0.019 & -\footnotemark[\value{mpfootnote}] & -\footnotemark[\value{mpfootnote}] & 0.021 & 0.024 & 0.017 \\
337.1 & 0.887 & 1.173 & 1.021 & 1.169 & 1.108 & 1.242 \\
344.6 & -\footnotemark[\value{mpfootnote}] & -\footnotemark[\value{mpfootnote}] & -\footnotemark[\value{mpfootnote}] & - & - & 0.000 \\
346.9 & 0.012 & 0.015 & -\footnotemark[\value{mpfootnote}] & 0.012 & 0.012 & 0.000 \\
350.0 & 0.014 & 0.013 & -\footnotemark[\value{mpfootnote}] & 0.014 & 0.014 & 0.012 \\
353.7 & 0.146 & 0.188 & 0.130 & 0.156 & 0.181 & 0.170 \\
357.7 & 0.707 & 0.889 & 0.799 & 0.931 & 0.882 & 0.889 \\
364.2 & -\footnotemark[\value{mpfootnote}] & -\footnotemark[\value{mpfootnote}] & -\footnotemark[\value{mpfootnote}]& - & - & 0.001 \\
367.2 & 0.009 & 0.012 & -\footnotemark[\value{mpfootnote}] & 0.010 & 0.010 & 0.005 \\
371.0 & 0.037 & 0.047 & -\footnotemark[\value{mpfootnote}] & 0.038 & 0.038 & 0.030 \\
375.5 & 0.150 & 0.187 & 0.238 & 0.155 & 0.180 & 0.160 \\
380.5 & 0.261 & 0.328 & 0.287 & 0.343 & 0.325 & 0.381 \\
385.8 & -\footnotemark[\value{mpfootnote}] & -\footnotemark[\value{mpfootnote}] & -\footnotemark[\value{mpfootnote}] & - & - & 0.001 \\
389.4 & 0.006 & -\footnotemark[\value{mpfootnote}] & -\footnotemark[\value{mpfootnote}] & 0.007 & 0.007 & 0.006 \\
391.4 & 0.281 & 0.454 & 0.302 & 0.315 & 0.315 & - \\
394.3 & 0.025 & 0.032 & 0.063 & 0.026 & 0.026 & 0.025 \\
399.8 & 0.090 & 0.119 & 0.129 & 0.097 & 0.113 & 0.085 \\
\hline
\end{tabular}
\end{center}
\end{minipage}
\end{table}
\begin{table}[t]
\renewcommand{\thefootnote}{\thempfootnote}
\begin{minipage}{\linewidth}
\begin{center}
\begin{tabular}{|c||c|c|c||c|c|c|}\hline
\multicolumn{7}{|c|}{\emph{continuation Table~\ref{tab:yield}}} \\ \hline
 & \multicolumn{3}{c||}{$FY_\lambda^{s.l.}~(\frac{{\rm photons}}{{\rm m}})$ measurements from} 
& \multicolumn{3}{c|}{$FY_\lambda^{s.l.}~(\frac{{\rm photons}}{{\rm m}})$ calculations with} \\
Wave- & Bunner & Davidson \& & Nagano & $\varepsilon_{\lambda,Bunner}^0$, & $\varepsilon_{\lambda,Bunner}^0$, 
& $\varepsilon_{\lambda,G.-F.}^0$, \\
length & \cite{bunner} & O'Neil~\cite{DO} & et al.~\cite{nagano}  & Table & Table & $\sigma_{Nx,Morozov}$, \\
$\lambda$ (nm) & & & & \ref{tab:deact}$_{Bunner}$ & \ref{tab:deact}$_{Morozov}$ & $\tau_{0,Gilmore}$ \\ \hline \hline
 \multicolumn{7}{|l|}{sum of $\lambda$ = (300-400) nm} \\
 & 3.001 & 3.490\addtocounter{mpfootnote}{+2}\footnotemark[\value{mpfootnote}] & 3.698 & 3.672 & 3.653 & 3.438\footnote{Without 1N band system.} \\ \hline
 \multicolumn{7}{|l|}{sum of all \emph{Nagano}-wavelengths} \\
 & 2.798 & 3.405\addtocounter{mpfootnote}{-1}\footnotemark[\value{mpfootnote}] & 3.698 & 3.460 & 3.438 & 3.283\addtocounter{mpfootnote}{+1}\footnotemark[\value{mpfootnote}] \\ \hline
 \multicolumn{7}{|l|}{sum of all \emph{Nagano}-wavelengths above 320~nm} \\
 & 2.574 & 3.405 & 3.149 & 3.221 & 3.159 & 2.957\footnotemark[\value{mpfootnote}] \\ \hline 
 \multicolumn{7}{|l|}{sum of all \emph{Nagano}-wavelengths above 320~nm and without 391.4~nm} \\
 & 2.293 & 2.951 & 2.847 & 2.906 & 2.844 & 2.957 \\  
\hline
\end{tabular}
\end{center}
\end{minipage}
\end{table}
and illustrated in Fig.~\ref{fig:spec_yield}.

\begin{figure}[htbp]
\centering\epsfig{file=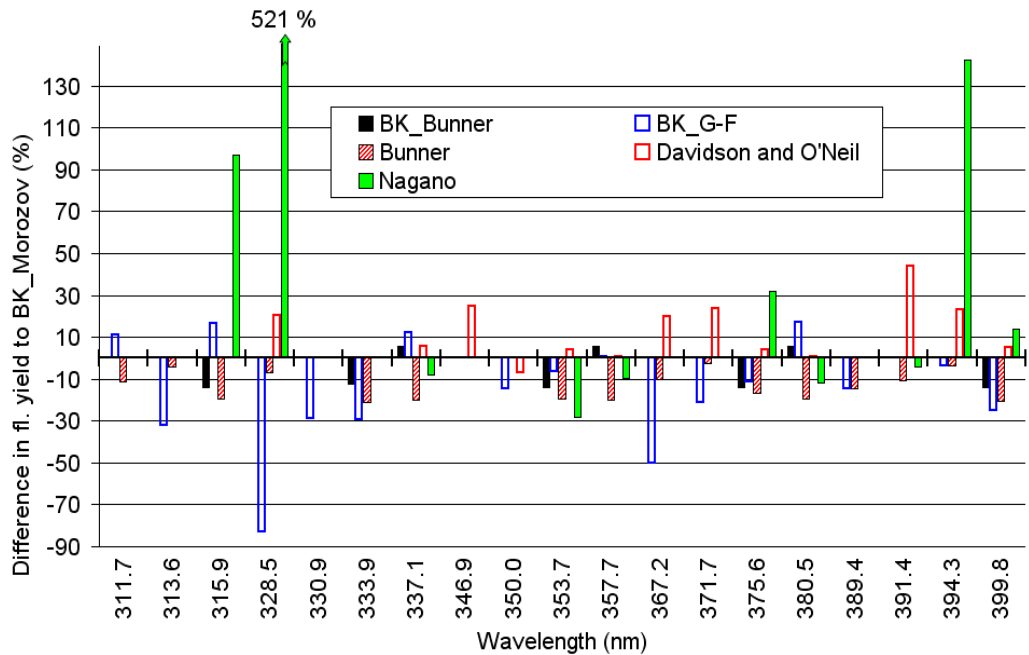,width=1.\linewidth}
\caption{Relative comparison of 19 band systems of the calculation
preferred in this article with measurements and further calculations. The absolute fluorescence yield of these
contributions can be seen in Fig.~\ref{fig:spec_yield} with the same marking. \label{fig:relVgl_autoren}}
\end{figure}

The total fluorescence yield reported by Bunner directly in \cite{bunner} is much lower than the  
other measurements and calculations. The total value from Davidson and O'Neil is higher by 6.8\% for
wavelengths above 320~nm as compared to our model for the same wavelength region. The calculations 
shown here applying $\varepsilon_{\lambda,Bunner}^0$ reproduce the measured values from Nagano et al.~very 
accurately and the partly differing deactivation constants from Bunner and Morozov et al.~do not affect 
the final result much. However, this holds only for the comparison of the whole wavelength region 
between 300 and 400~nm. One difficulty in the measurements is the treatment of interference filters 
which have a bandwidth of about 10~nm~\cite{nagano}. The 10 contributions of Nagano et al.~are given 
after subtracting additional contributions by smaller emissions within one filter region.
Thus, for a direct comparison, one has to take into account only the 10 wavelengths reported in \cite{nagano}
and in this case, the calculations with $\varepsilon_{\lambda,Bunner}^0$ differ by approximately -7\%. 
For a detailed comparison of each individual band system, see Fig.~\ref{fig:relVgl_autoren}.

The measurements from Nagano et al.~show two contributions which are considerably larger than all
other data for those band systems at 315.9 and 394.3~nm. The largest relative difference occurs
at the wavelength 328.5~nm where the value measured by Nagano et al.~is higher by 512\% compared
to the preferred calculation presented here. However, the absolute contribution of this band emission is only of 
minor importance for the entire wavelength region between 300 and 400~nm. 

The calculations based on the Einstein coefficients given by Gilmore et al.~lead to typically 20\% - 30\% 
lower fluorescence yield than our preferred calculation. Interestingly, the band emission at 328.5~nm, 
which is very bright in the measurements by Nagano et al., is much lower in the calculations using 
Gilmore et al.~combined with Fons et al.~data than in our model. These uncertainties might be caused by possible 
contributions from the 1N system, see Fig.~\ref{fig:spec_yield} or Tab.~\ref{tab:fleff_1N}.

Hirsh et al.~\cite{hirsh} have performed measurements for the 1N (0-0) band system of nitrogen.
They found a value for the fluorescence efficiency $\varepsilon_{391.4~nm}^0$ of 0.475\% which is 
considerably higher than the value given by Bunner, see Table~\ref{tab:fleff}. Additionally, the 
collisional cross section of nitrogen with nitrogen and nitrogen with oxygen have been investigated
\cite{hirsh}. The values are $\sigma_{NN}$ = 6.5$\times 10^{-19}$~m$^2$ and 
$\sigma_{NO}$ = 10.9$\times 10^{-19}$~m$^2$. Calculating the
fluorescence yield for a 0.85~MeV electron with these parameters, the value at sea level in the 
US Standard Atmosphere amounts to 0.377~$\frac{{\rm photons}}{{\rm m}}$. A comparison of this 
number with the entries in Tab.~\ref{tab:yield} shows that $FY_{391.4~nm}$ measured by Hirsh 
et al.~is larger than our calculation by 20\%, larger than measurements from Nagano et al.~by 25\%, and 
larger than measurements published by Bunner even by 34\%. Only measurements performed by Davidson and O'Neil
result in a 17\% higher $FY_{391.4~nm}$ than Hirsh et al.

Concluding, it can be stated that the calculations shown here provide a reasonable way of describing
fluorescence emission in air while allowing for varying atmospheric conditions. This procedure can
easily be implemented into air shower reconstruction programs. The overall agreement in 
the wavelength region between 300 and 400~nm with some measurements is already satisfying. 
However detailed, spectrally resolved comparisons reveal uncertainties
in measurements and the understanding of the processes in air. Further investigations are
necessary, as both atmospheric transmission and EAS detection are wavelength dependent. 
Also the dependence on altitude is different for the emission bands. Thus, for reconstructing 
the fluorescence emission of an EAS in the atmosphere from the measured photons, all
these processes must be understood. It must be stressed that, 
for the EAS experiments, the uncertainties in fluorescence yield are directly linked to the 
uncertainties in the primary energy of cosmic rays.

It should also be mentioned that further fluorescence yield measurements, however without
spectral resolution, can be found in literature. 
Kakimoto et al.~provide a formula for calculating the fluorescence yield between 300 and 400~nm, which 
gives at sea level 3.275~$\frac{{\rm photons}}{{\rm m}}$~\cite{kakimoto}. This value is 
smaller by 
10.3\% compared to our preferred calculation. The HiRes Collaboration uses a value of about 
5~$\frac{{\rm photons}}{{\rm m}}$ per charged particle in an air shower~\cite{hires2}. 
For these charged particles, an average energy deposit of 0.22 MeV/kg m$^{-2}$ is assumed~\cite{bunner}, 
which leads to a corresponding fluorescence yield at s.l.~of 3.811~$\frac{{\rm photons}}{{\rm m}}$ for a 
0.85~MeV electron. Assuming that the HiRes value refers to 5~km~a.s.l., one would obtain 
3.6 - 3.7~$\frac{{\rm photons}}{{\rm m}}$ at s.l.

\section{Dependence on Atmospheric Conditions\label{sec.althumdep}}

\subsection{Altitude Dependence}

\begin{figure}[htbp]
\centering\epsfig{file=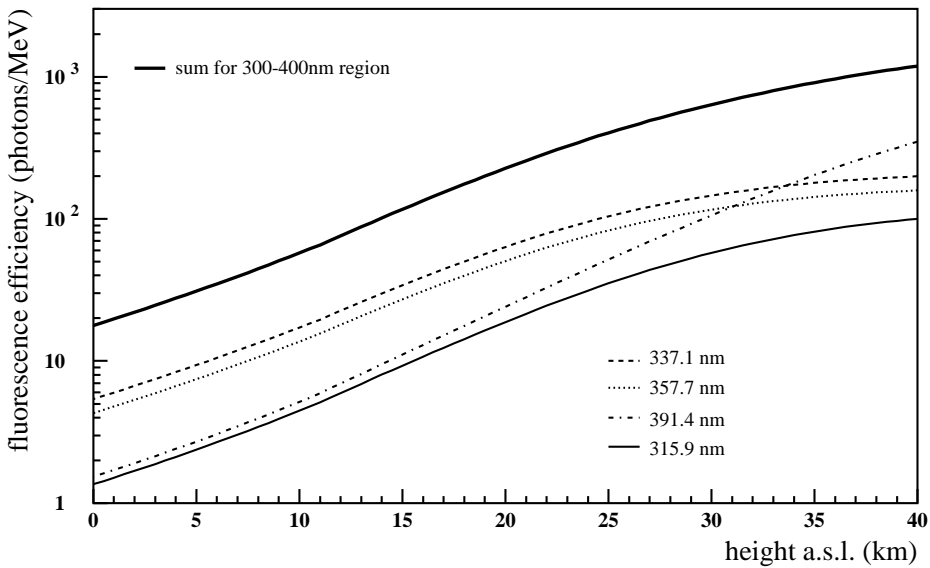, width=0.9\linewidth}
\caption{Fluorescence efficiency profiles for different wavelengths in the US-StdA.\label{fig:altdep}}
\end{figure}

Firstly, the altitude dependence of the fluorescence efficiency $\varepsilon_{E_{dep}}$ in units of 
photons per MeV of deposited energy, as described in Section~\ref{sec.comp}, will be shown. Equations 
(\ref{eq.ppprime}) and (\ref{eq.ppprimelong}) contain the altitude dependent variables $p$ and $T$.
Combined with the deactivation constants, Tab.~\ref{tab:deact}, a different altitude dependence for
the 1N (391.4~nm) and 2P system (all other wavelengths) is expected, which is visualized in 
Fig.~\ref{fig:altdep}.
Within the 2P system of nitrogen, there are no differences in the altitude dependence for all
emission bands belonging to the same excited state $\nu^\prime$ (e.g.~337.1~nm and 357.7~nm) and 
only small variations are between bands of different $\nu^\prime$ states (compare e.g.~315.9~nm from
$\nu^\prime = 1$ and 337.1~nm from $\nu^\prime = 0$). With increasing altitude, the efficiency becomes
larger due to lower rates of collisional quenching. This increase is largest for the 391.4~nm band. 
At sea level its contribution to the total spectrum amounts to 8.6\%, at 20~km~a.s.l.~it is already
10.7\%, and at 30~km~a.s.l.~16.8\%. 

\begin{figure}[tbp]
\centering\epsfig{file=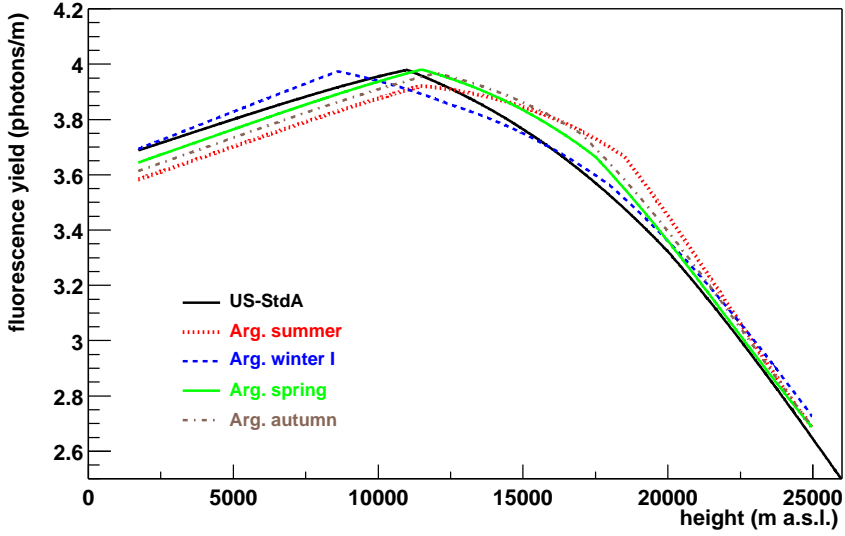, width=0.9\linewidth}
\caption{Fluorescence yield profiles for a 0.85~MeV electron with vertical incidence in 
the US Standard atmosphere and measured Argentine atmospheres as given in \cite{keilhauer04}. 
The given yield is a sum of all emitted
photons between 300 and 400~nm calculated as described in Sec.~\ref{sec.comp} with 
$\varepsilon_{\lambda,Bunner}^0$ and Table~\ref{tab:deact}$_{Morozov}$.\label{fig:flyield_alt}}
\end{figure}

However, regarding EAS, the rate of emitted photons per meter
traversed matter of the EAS is the observed variable which is the fluorescence yield $FY_\lambda$, 
Eq.~(\ref{eq.FY}). Via the altitude-dependent air density, $\rho_{air}$, the number of excitable
nitrogen molecules and quenching partners are considered. For simplicity, $FY_\lambda$ 
vs.~altitude is shown in Fig.~\ref{fig:flyield_alt} for a 0.85~MeV electron.
The most relevant altitude range for EAS is between ground and about 13~km~a.s.l. For EAS with 
energies of about 10$^{19}$~eV, the shower reaches its maximum between 2 and 8~km~a.s.l.~depending
on the type of the primary particle and the inclination angle of the EAS. 

E.g.~for the Auger experiment, the field of view of a telescope covers an altitude range between 
0.7~km and 12.5~km above the altitude of the Pierre Auger Observatory, 1.4~km a.s.l., at a distance 
of 20~km. The fluorescence yield plotted in Fig.~\ref{fig:flyield_alt} is a sum of all 
emitted photons between 300 and 400~nm calculated as described in 
Sec.~\ref{sec.comp} with $\varepsilon_{\lambda,Bunner}^0$ and Table~\ref{tab:deact}$_{Morozov}$.
Additionally to the altitude dependence, also the seasonal dependence for actual atmospheres as 
obtained at the southern site of the Pierre Auger Observatory \cite{keilhauer04} can be seen in 
Fig.~\ref{fig:flyield_alt}. From
ground level to altitudes around 10~km, the fluorescence yield increases slowly. Above 10~km, the yield decreases
disclosing the sensitivity to temperature and pressure variations. During winter I, the lower
temperatures compared to the other atmospheric models below 9~km~a.s.l.~induce a higher
fluorescence yield. Up to 17~km, the temperatures are comparatively high leading to a reduced
fluorescence yield. During spring, summer, and autumn the temperatures are higher than in the
US-StdA, therefore the fluorescence yield is decreased mostly in summer. Above
15~km~a.s.l., the very low temperatures during summer result in a very high emission. The differences
of $FY_\lambda$ for the Argentine seasons compared to the US-StdA are well below
$\pm$5\%. At the altitude of the Auger detectors, 1.4~km~a.s.l., the increase in fluorescence 
yield during winter~I is negligible,
however the decrease in summer amounts to 2.9\%. At $\approx$ 8.5~km, the differences of summer and 
winter~I to the US-StdA are of the same size but with opposite signs. In winter~I, $FY_\lambda$
is 1.5\% higher than in the US-StdA, and in summer 2.2\% lower. More than +4\% difference from 
Argentine summer to the US-StdA emerges above 16.5~km~a.s.l. Similar seasonal variations in 
$FY_\lambda$ are also valid for other EAS experiments since similar atmospheric conditions have 
been found at different places~\cite{keilhauer04,wilczynska}. 

\begin{figure}[htbp]
\centering\epsfig{file=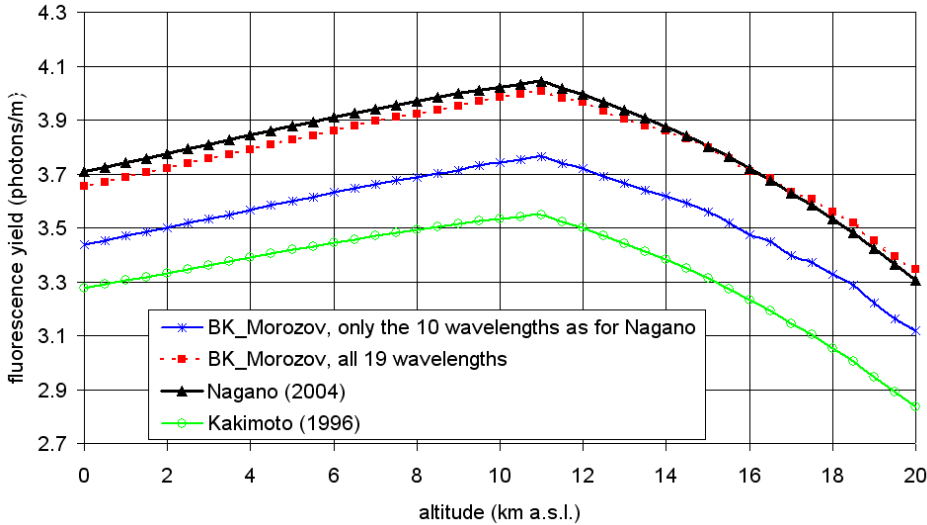, width=0.9\linewidth}
\caption{Fluorescence yield profiles for a 0.85~MeV electron in the US-StdA.
Comparison of the altitude dependence calculated by the described method with two further
parameterizations.\label{fig:yield_hoeheUS_Nagano_KUl_Kak}}
\end{figure}

The calculated altitude dependence can be compared with parameterizations given by authors from
fluorescence emission experiments. The functional forms of these parametrizations are
inspired by the same equations as introduced in Sec.~\ref{sec.mathdesc} \cite{nagano,kakimoto}:
\begin{eqnarray}
FY_\lambda^{\textrm{\scriptsize{\cite{nagano}}}}&=&\frac{dE}{dX}\cdot\biggl(\frac{A_\lambda\rho}
{1+\rho B_\lambda \sqrt{T}}\biggr), \\
FY_{300-400~nm}^{\textrm{\scriptsize{\cite{kakimoto}}}}&=&\frac{dE}{dX}\cdot\rho\biggl(\frac{A_1}
{1+\rho B_1 \sqrt{T}}+\frac{A_2}{1+\rho B_2 \sqrt{T}}\biggr).
\end{eqnarray}  
While Nagano et al.~\cite{nagano} list $A$ and $B$ parameters for each of their 10 wavelengths between
300 and 400~nm, Kakimoto et al.~\cite{kakimoto} just provide one set of parameters $A_{1,2}$ and 
$B_{1,2}$ for the total fluorescence yield between 300 and 400~nm.
Both approaches predict similar height dependences, see Fig.~\ref{fig:yield_hoeheUS_Nagano_KUl_Kak}.
To work out the difference due to the altitude dependence, the profiles can be shifted so that
all curves start with the same value at sea level. Then the parameterization by Nagano 
et al.~agrees very well with the calculation introduced in this paper. Up to 14~km, the discrepancy is 
below 1\%, increasing up to 2.9\% at 20~km~a.s.l.
The simplified parameterization given by Kakimoto et al.~disagrees already above 6.5~km with the 
calculations shown here by more than 1\%. The difference increases up to 4\% at 20~km~a.s.l.

\subsection{Humidity Dependence}

All calculations and measurements shown above are based on dry air conditions. However, in actual atmospheric
conditions, there is sometimes a considerable fraction of water vapor. Thus, the effect of quenching
due to water vapor has to be investigated. Fluorescence emission by water vapor is not expected.

Our first calculations are based on Eq.~(\ref{eq.ppprimelong}) in which an additional term
is inserted to account for the collisions between nitrogen and water vapor molecules. The
experimental determination of collisional cross sections between nitrogen and water vapor,
which are needed in that equation,  
is very difficult. Two experiments have recently begun to investigate the effect of water vapor
\cite{ulrich,nagano2}. 

The effect of quenching due to water vapor has been studied in our preferred calculation for the 
337.1~nm emission band. 
Applying the parameters from Tab.~\ref{tab:deact} by Morozov et al.~and assuming 100\% relative humidity, the
emission at sea level is reduced by approximately 20\%, at 4~km~a.s.l.~by roughly 5\%, and at 
8~km~a.s.l.~just by 0.3\%. Since fluorescence telescopes typically operate only during ``good weather''
periods, this decrease in fluorescence yield should be considered as an upper limit. For realistic atmospheric 
conditions, a reduction of about 5 to 10\% near ground and 1 - 3\% at 4~km~a.s.l.~can be expected. 

\section{Summary and Conclusion}
\label{sec.summary}

EAS experiments applying the fluorescence technique measure the light emission in air induced 
by charged particles, mainly electrons and positrons. The detected
light track is converted into a longitudinal shower profile and finally to the total energy of the
primary particle of the EAS. Therefore, the fluorescence light yield has to be known precisely 
including spectral resolution and dependent of atmospheric conditions.

The results on fluorescence yield which can be found in literature differ considerably. 
Most important are the fluorescence efficiency of the contributing
band systems of nitrogen, but also the radiative life times and the collisional cross sections
of nitrogen with nitrogen and nitrogen with oxygen have to be known. Up to now, a thorough understanding
of the energy-dependent excitation processes of the different nitrogen states is missing. First studies
can be found in Blanco and Arqueros~\cite{blanco}.

In this article, an atmosphere-dependent description of the fluorescence light emission in air has been presented.
The different contributions of the 2P and 1N band systems of nitrogen have been calculated in detail. The
calculations are based on several parameter sets and have been compared with fluorescence yield measurements 
performed by several authors. The calculations reproduce some results of measurements well, while 
other data are off by more than 10\% regarding the total yield between 300 and 400~nm. The differences for 
individual emission bands are much larger. 

The variation of the fluorescence yield with changing atmospheric
conditions has only been studied by a few authors. Generally, it is assumed that the main reduction of 
light emission is due to collisional quenching. The calculation of altitude-dependent profiles of $FY_\lambda$ 
presented here agree within 4\% with parameterizations of measurements. 

Using the calculation and parameter set preferred in the article, a prediction of the influence 
of water vapor has been made. For realistic
atmospheric conditions, an effect of about 5 to 10\% near ground and less than 
3\% at altitudes around 4~km~a.s.l.~can be expected.
Only lately experimental studies of the quenching rate of water vapor have been begun.

\section*{Acknowledgment}
One of the authors (BK) is supported by the German Research Foundation (DFG) under contract No.~KE 1151/1-1.

\end{document}